\documentclass[prb,twocolumn,showpacs,preprintnumbers,amsmath,amssymb]{revtex4}
\usepackage{amsfonts}
\usepackage{graphicx}
\usepackage{dcolumn}
\usepackage{bm}

\begin{document}

\title{Quantum phase transition in the multimode Dicke model}
\date{\today }
\author{Denis Tolkunov\footnote{Electronic address: Tolkunov@clarkson.edu} and Dmitry Solenov\footnote{Electronic address:
Solenov@clarkson.edu}} \affiliation{Department of Physics,
Clarkson University, Potsdam, New York 13699--5820}
\date{\today}

\begin{abstract}
An investigation of the quantum phase transition in both discrete
and continuum field Dicke models is presented. A series of
anticrossing features following the criticality is revealed in the
band of the field modes. Critical exponents are calculated. We
investigate the properties of a pairwise entanglement measured by
a concurrence and obtain analytical results in the thermodynamic
limit.
\end{abstract}


\maketitle

%
\textit{Introduction.} An ensemble of two-level quantum systems,
each coupled to the common radiation field modes, the Dicke
model,\cite{Dicke} was introduced initially to investigate
superradiant emission---a dramatic increase in the rate of
coherent spontaneous emission of an ensemble of atoms. The model
has found a variety of applications in quantum optics, nanoscale
solid-state physics, and quantum information theory. The critical
behavior of this model was first discussed in
Refs.~\onlinecite{HeppLieb, Hioe, Duncan}, where the single-mode
Dicke model has been shown to admit a second-order classical phase
transition. An investigation of the thermodynamic properties of
the multimode model of superradiance was done in
Refs.~\onlinecite{HeppLieb_multimode} and \onlinecite{Emeljanov}
where the conditions for the thermodynamic instability were
derived. In Refs.~\onlinecite{Rzazewski} and
\onlinecite{Rzazewski2} Rz\c{a}\.zewski \textit{et al.} have
demonstrated that the presence of the $A^2$ term in the
interaction leads to the disappearance of the classical phase
transition in the single-mode model. The multimode case has been
addressed in Refs.~\onlinecite{Thompson1} and
\onlinecite{Thompson2}.

The quantum, or zero-temperature, phase transition reveals itself
by the presence of nonanalyticity in the {\em ground-state energy}
of a quantum system as some parameter of the system---e.g., the
strength of the interaction---is varied.\cite{Sachdev} In addition
the energy gap---e.g., between the ground and first excited
states---vanishes at the critical point as a power-law function.
This kind of phase transition is driven by quantum fluctuations
and related to qualitative changes in the {\em ground state} of
the system. A single-mode Dicke model \cite{Dicke} has been shown
to be one of the examples of such
behavior.\cite{Wang-Hioe,Hillery-Mlodinow,Emary-Brandes} Recently
Emary and Brandes\cite{Emary-Brandes} have considered a
single-mode Dicke model with no rotating-wave approximation and
demonstrated that at the quantum phase transition the system
changes from being quasi-integrable to quantum chaotic. In
Refs.~\onlinecite{Reslen, Vidal, Plastina} the universality
properties of the single-mode Dicke model have been addressed; it
was demonstrated that the finite-size scaling hypothesis
\cite{Privman} works well for the single-mode Dicke model---it
pertains to the same universality class as the infinite-range
anisotropic \textit{XY} model in a transverse field. The divergent
correlation length is typical for the quantum as well as classical
transition at criticality. Quantum systems, however, acquire
additional correlations that do not have a classical counterpart.
These are entanglement. In connection with the quantum phase
transition in the single-mode superradiance model, it has been
studied in Refs.~\onlinecite{Emary-QPT-Concurrence, Vidal}, and
\onlinecite{Plastina}.

In the present paper we study a general model
\cite{HeppLieb_multimode, Emeljanov, Thompson1, Thompson2} in
which the atomic system is coupled to an arbitrary (discrete or
continuous) set of radiation modes with arbitrary coupling
constants. Many recent publications have emphasized the
achievements in creating a nearly single-mode environment, such
as, for instance, high-\textit{Q} cavities.\cite{QCavities} One
should, however, note that the cavities used currently have
essentially a finite quality factor. In fact it is rather a tough
problem in many circumstances to create an environment with a
single field mode.\cite{QCavities} Therefore the multimode
generalization provides a more realistic scenario. In addition,
significant progress in solid-state photonic
crystals,\cite{PhCrGap} systems of artificial
atoms,\cite{Microdisk} and ion-trap
structures\cite{ion-trap-cavity} have allowed experimental
investigation of a rich variety of atom-field coupling regimes, in
which case the multimode Dicke model may be an exciting target.

%
\textit{Multimode Dicke model.} We consider a system of $M$ atoms
distributed over some finite volume. Each atom is regarded within
the two-level approximation. The atoms are identical, having the
same transition frequency $\omega_0$, and coupled to the
electromagnetic radiation field via a dipole interaction with the
coupling constants $g^j_k$ for the $j$th atom and $k$th mode. We
also assume that the system resides in a volume small enough such
that the electromagnetic field is nearly uniform within the atomic
distribution. Therefore the atom-field coupling constants depend
weakly on the atom's number---i.e., $g^j_k\approx g_k$. The
overall Hamiltonian of the model is
\begin{eqnarray}\nonumber \label{Eq:H}
H&=&\frac {\omega _0}{2}\sum_{j=1}^M\sigma _z^j + \sum_k\omega
_ka_k^{\dagger }a_k
\\&+&\frac
1{\sqrt{M}}\sum_{k}\sum_{j=1}^M g_k\left( a_k^{\dagger
}+a_k\right) \left( \sigma _{+}^j+\sigma _{-}^j\right).
\end{eqnarray}
We use units with $\hbar=1$. The atom-field coupling constants are
proportional to the inverse square root of the atomic distribution
volume, $V\sim M$. We explicitly show this $1/\sqrt{M}$ dependence
in the interaction term, whereas $g_k$ are finite and independent
of $M$. Note that Eq.(\ref{Eq:H}) has parity symmetry; i.e. it
commutes with $\Pi =\exp \left[ i\pi \left( \sum_ka_k^{\dagger
}a_k+\sum_j\sigma _z^j\right) \right]$.

The model given by Eq.(\ref{Eq:H}) incorporates chaotic dynamics
and is not integrable even for a single-mode field. However, it
becomes approachable in the thermodynamic limit
$M\rightarrow\infty$ for a large number of atoms. In this limit
the solution of the single-mode model was carried out in
Refs.~\onlinecite{Hillery-Mlodinow} and \onlinecite{Emary-Brandes}
where it was demonstrated that the system undergoes a quantum
phase transition. Below we present a solution to the multimode
model in a form convenient for further analysis of criticality and
entanglement.

%
\textit{Effective Hamiltonian.} Taking into account the structure
of the Hamiltonian (\ref{Eq:H}) we introduce the collective spin
operators\cite{Dicke} $J_{\pm}\equiv \sum_i\sigma _{\pm}^i$ and
$J_z \equiv \frac{1}{2}\sum_i\sigma _z^i$. The corresponding
square of the total spin operator, $J^2$, with eigenvalues
$j(j+1)$ commutes with $H$, and therefore $j=M/2$ is a good
quantum number. In our analysis we are interested in the situation
when all atoms are at the ground state before the transition, so
that the expectation of $J_z$, proportional to the magnetization
of the atomic system, is $-M/2$. Using the Holstein-Primakoff
transformation \cite{Holstein-Primakoff}
\begin{eqnarray}\nonumber \label{Eq:HP-relations}
J_{+}&=&b^{\dagger }\sqrt{M-b^{\dagger }b},\,\,\,
J_{-}=J_{+}^{\dagger},
\\J_z&=&b^{\dagger }b-\frac M2,
\end{eqnarray}
where $b$ and $b^{\dagger}$ are the usual bosonic operators, one
can expand the Hamiltonian (\ref{Eq:H}) in powers of $1/M$.
Neglecting the terms of order ${\cal O}(M^{-s})$, $s\geq 1/2$, we
obtain the effective Hamiltonian in the form
\begin{equation}\label{Eq:H-eff-Before-QFT}
H_{eff}\!\!=\!\omega _0b^{\dagger }b+\sum_k\omega _ka_k^{\dagger
}a_k+\!\sum_kg_k\big( a_k^{\dagger }+a_k\big)\! \left( b^{\dagger
}+b\right) -\omega _0j.
\end{equation}
When the field modes are distributed continuously and separated
from $\omega_0$ by a gap, the first term represents a localized
impurity. On the other hand, placing $\omega_0$ within the field
spectrum leads to the situation when confinement of the
corresponding state may be defined only for a certain lifetime,
represented by the imaginary part of the ``energy.'' In the
essentially discrete model of the field modes, $k=1,2,\ldots,N$,
which the single-mode case is an extreme example of, making
$\omega_0$ equal to one of the field modes energies $\omega_k$
does not change the situation dramatically. Nevertheless, the
diagonalization of Eq.(\ref{Eq:H-eff-Before-QFT}) can be carried
out in the discrete case, provided proper integrations are
introduced in the final equations when the continuous-model result
is sought.

The Hamiltonian (\ref{Eq:H-eff-Before-QFT}) is quadratic and can
be brought to the diagonal form $H^\prime=\sum_\alpha \varepsilon
_\alpha q_\alpha ^{\dagger }q_\alpha$ by the Bogolubov
transformation defining a new set of bosonic variables
\begin{equation}\label{Eq:Bogolubov}
a_\alpha\!\!=\!\!\sum_\beta \left(A_{\alpha \beta }q_\beta
\!+\!\overline{A}_{\alpha \beta }q_\beta^{\dagger }\right),\,\,
q_\beta\!\!=\!\!\sum_\alpha \left(B_{\beta \alpha }a_\alpha
+\overline{B}_{\beta \alpha }a_\alpha^{\dagger }\right),
\end{equation}
where $A_{\alpha \beta }$, $\overline{A}_{\alpha \beta }$,
$B_{\beta \alpha }$, and $\overline{B}_{\beta \alpha }$ are
complex amplitudes representing the direct and inverse
transformations. For convenience, we introduced Greek indexes
$\alpha,\beta,\dots$, such that $a_{\alpha=0}^{(\dagger)}\equiv
b^{(\dagger)}$ and $a_{\alpha=k}^{(\dagger)}\equiv
a_k^{(\dagger)}$, assuming that $k$, as a discrete index, counts
the modes beginning with ``1,'' as suggested earlier.

The easiest way to proceed with the diagonalization is to use the
commutation approach. One equates the coefficient before the
creation (annihilation) operators in
$\big[a_\alpha,H_{eff}^{\prime }\big] =\big[
a_\alpha,H_{eff}\big]$ as soon as these commutators are evaluated
using the direct (inverse) transformation (\ref{Eq:Bogolubov}).
This procedure leads to the system of equations
\begin{eqnarray}\nonumber\label{Eq:System-From-Direct-Transformation}
&&\omega _\alpha \!A_{\alpha \beta }^{+}\!+\!2\delta _{\alpha
k}g_kA_{0\beta }^{+}\!+\!2\delta_{\alpha 0}\!\sum_kg_kA_{k\beta
}^{+}\!=\!\varepsilon_\beta A_{\alpha \beta }^{-},\\
&&\omega _\alpha A_{\alpha \beta }^{-}\!=\!\varepsilon_\beta
A_{\alpha \beta }^{+},
\end{eqnarray}
where $A_{\alpha \beta }^{+}=A_{\alpha \beta
}+\overline{A}_{\alpha \beta }$ and $A_{\alpha \beta
}^{-}=A_{\alpha \beta }-\overline{A}_{\alpha \beta }$. The system
becomes complete provided the bosonic commutation relations
$\big[q_\alpha,q_\beta^\dagger \big]=\delta_{\alpha\beta}$ and
$\big[a_\alpha,a_\beta^\dagger \big]=\delta_{\alpha\beta}$ are
taken into account. One can also benefit significantly from the
fact that $B_{\alpha\beta}=A^*_{\beta\alpha}$ and
$\overline{B}_{\alpha\beta}=-\overline{A}_{\beta\alpha}$. This is
easily obtained playing around with the commutations of $a_\alpha$
and $q_\alpha$ with Eq.(\ref{Eq:Bogolubov}). The complete solution
of the system (\ref{Eq:System-From-Direct-Transformation}) is
\begin{eqnarray}\nonumber\label{Eq:Solution-To-Direct-Transformation}
&&A_{0\beta }=\frac{\omega _0+\varepsilon _\beta }{\sqrt {4\omega
_0\varepsilon _\beta} }\Bigg[ 1+\sum_k\frac{4g_k^2\omega _0\omega
_k}{\big( \omega _k^2-\varepsilon _\beta ^2\big) ^2}\Bigg]
^{-1/2},
\\\nonumber
&&A_{k\beta }=-\frac{g_k}{\omega _k-\varepsilon _\beta }\left(
\frac{\omega _0-\varepsilon _\beta }{\omega _0+\varepsilon _\beta
}+1\right) A_{0\beta },
\\
&&\overline{A}_{\alpha \beta }=\frac{\omega _\alpha -\varepsilon
_\beta }{ \omega _\alpha +\varepsilon _\beta }A_{\alpha \beta }.
\end{eqnarray}
The inverse transformation is obtained recalling that
$B_{\alpha\beta}=A^*_{\beta\alpha}$ and
$\overline{B}_{\alpha\beta}=-\overline{A}_{\beta\alpha}$.

%
\textit{Below the critical point.} For $g_k>0$, the spectrum of
the model is given by the solutions to the equation
\begin{equation}\label{Eq:E-Discrete-Before-QFT}
4\omega _0\sum_{k=1}^N\frac{g_k^2\omega _k}{\omega
_k^2-\varepsilon ^2} +\varepsilon ^2-\omega _0^2=0.
\end{equation}
For the single-mode case ($N=1$) the solution coincides with one
derived in Ref.~\onlinecite{Emary-Brandes}. In general, for any
$N$, the above equation cannot be solved analytically. However, an
important property can be revealed from a mere analysis of
Eq.(\ref{Eq:E-Discrete-Before-QFT}). It can be noticed that the
excitation energies $\varepsilon$ are real only for
$\gamma<\gamma_c$, where $\gamma =\sum_kg_k^2/\omega _k$ and
$\gamma_c=\omega_0/4$. This defines \emph{the critical point},
with the relation between the coupling constants
$4\sum_kg_k^2/\omega _k=\omega _0$. This critical value of the
interaction defines the quantum phase transition point, separating
normal and superradiant phases. After the critical point
$\gamma>\gamma_c$, the energies are found by properly displacing
the bosonic modes, as will be shown shortly. For the moment let us
discuss the continuum model corresponding to
Eq.(\ref{Eq:E-Discrete-Before-QFT}).

In the case of the continuum there is a difference between the
bosons created by the impurity operators $b^{\dagger }$ and the
field (bath) operators $a_k^{\dagger }$. The bath energies are not
affected significantly; i.e., one can reasonably assume that the
impurity does not change the spectrum of the bath, $\omega _k$. At
the same time, the bath does dress the impurity state. Further we
assume that the impurity is localized and the corresponding
dressed energy is outside of the field band.

For the continuous field spectrum model, in
Eq.(\ref{Eq:E-Discrete-Before-QFT}) one replaces
$\sum^{N\rightarrow \infty}_{k=1}\rightarrow \int^\infty_0 d\omega
\mathcal{D}(\omega)$ and $g_k\rightarrow g(\omega)$, where
$\mathcal{D}(\omega)$ is the density of the bath modes. As an
instructive example, we model the density of modes and the
coupling constants as $\mathcal{D}_\omega g^2\left( \omega \right)
=\alpha \omega^n \theta(\omega-\Omega_1)\theta(\Omega_2-\omega)$
with ${n=1}$, which corresponds to the well-known Ohmic
case.\cite{Leggett} The width of the band is, then, $\Delta \equiv
\Omega _2-\Omega _1$. This choice is for convenience only. It does
not affect the qualitative structure of the result. Finally, the
impurity level is given by the transcendental equation
\begin{equation}\label{Eq:E-Continuous-Before-QFT}
2\alpha\varepsilon \omega _0\ln \frac{\Omega _1+\varepsilon
}{\Omega _2+\varepsilon }\frac{\Omega _2-\varepsilon }{\Omega
_1-\varepsilon }+\varepsilon ^2-\omega _0^2+4\alpha\omega _0\Delta
=0\,.
\end{equation}
Here the parameter $\alpha$ represents the strength of the
coupling and can be varied. The critical point, as follows from
Eq.(\ref{Eq:E-Continuous-Before-QFT}), is at $\alpha _c=\omega
_0/4\Delta$. Note that in Eq.(\ref{Eq:E-Continuous-Before-QFT})
the bandwidth has to be finite for the quantum phase transition to
appear---i.e., for $\alpha_c>0$. In most physical systems,
however, the effective width of the band, $\Delta_\textrm{eff}$,
is finite due to the natural cutoff that comes from the density of
modes, $\mathcal{D}(\omega)$, or the coupling constants
$g(\omega)$---i.e., $\int^\infty_0 d\omega f(\omega)\sim
\Delta_\textrm{eff}$ where $f(\omega)$ is some cutoff function. As
a result, one simply has $\alpha_c=\omega_0/4\Delta_\textrm{eff}$.

%
\textit{Above the critical point.} The quantum phase transition
results in a qualitative change in the structure of correlations
in the system's ground state.\cite{Sachdev} As the interaction
strength $g$ ($\gamma$) or $\alpha$ increases beyond the critical
point the parity symmetry of the system is broken and all the
oscillators in Eqs. (\ref{Eq:H}) and (\ref{Eq:HP-relations})
obtain new equilibrium positions; i.e., the operators $a_\alpha$
and $a^\dag_\alpha$ acquire \textit{c}-number shifts.

To get the effective Hamiltonian above the critical point we
displace all the bosonic modes in Eqs. (\ref{Eq:H}) and
(\ref{Eq:HP-relations}) by $a_k\rightarrow a_k+\sqrt{j}\alpha _k$
and $b\rightarrow b+\sqrt{j}\beta$, where $\alpha _k$ and $\beta$
are some complex constants and the factor $\sqrt{j}$ is for
convenience of notation. One can easily show, following the logic
of Ref.~\onlinecite{Hillery-Mlodinow}, that $\alpha _k$ and
$\beta$ are ${\cal O}(1)$ quantities,
\begin{equation} \label{Eq:H-eff-After-QFT}
\beta=\sqrt{1-\gamma_0/\gamma}, \,\,\,\,\,\,
\alpha_k=-\frac{g_k}{\omega _k}\gamma \beta\sqrt{2}\,.
\end{equation}
The effective Hamiltonian above the phase transition becomes
\begin{eqnarray}\nonumber
\widetilde{H}_{eff}\!\!\!&=&\!\!\sum_k\omega_ka_k^{\dagger}a_k+\widetilde{\omega
}_0b^{\dagger}b+\sum_k\widetilde{g}_k\big( a_k^{\dagger
}+a_k\big)\! \big(b+b^{\dagger }\big)
\\ \label{Eq:H-eff-After-QFT}
&+&\!\zeta\left(b^2+b^{\dagger 2}\right)\!+\xi ,
\end{eqnarray}
where
\begin{eqnarray}\label{Eq:Constants-To-H-eff-After-QFT}\nonumber
\widetilde{\omega }_0\!\!&=&\!\!\frac{\omega _0\left( 5+2\eta
+\eta ^2\right) }{4\eta \left( 1+\eta \right) },
\,\,\,\,\,\,\,\widetilde{g}_k=\eta \sqrt{\frac 2{1+\eta }}g_k,
\\\nonumber
\zeta \!\!&=&\!\!\frac{\omega _0\left( 1-\eta
\right) \left( 3+\eta \right) }{8\eta \left( 1+\eta \right) },
\\
\xi \!\!&=&\!\!\frac{\omega _0\left( 1-\eta \right) ^2}{8\eta
\left( 1+\eta \right) }-j \frac{\omega _0\left( 1+\eta ^2\right)
}{2\eta }\,.
\end{eqnarray}
Here we also introduced $\eta =\gamma_c/\gamma $, recalling the
earlier definitions $\gamma =\sum_kg_k^2/\omega _k$ and $\gamma_c
=\omega_0/4$. At the critical point, when $\eta =1$ this effective
Hamiltonian coincides with Eq.(\ref{Eq:H-eff-Before-QFT}).

The diagonalization procedure here is similar to the one below the
critical point. One makes the linear transformation of the form
(\ref{Eq:Bogolubov}) which leads to
$\widetilde{H}_{eff}^\prime=\sum_\alpha \varepsilon _\alpha
q_\alpha ^{\dagger }q_\alpha$. The spectrum above the critical
point is defined by the solution to the equation
\begin{equation}\label{Eq:E-Discrete-After-QFT}
\frac{\omega_0^2}{\gamma}\sum\limits_{k=1}^N\frac{g_k^2\omega
_k}{\omega _k^2-\varepsilon ^2}+\varepsilon^2-16\,\gamma^2=0.
\end{equation}
At the critical point, when $\gamma=\gamma_c$($=\omega_0/4$), this
equation coincides with Eq.(\ref{Eq:E-Discrete-Before-QFT}). The
coefficients of the transformation (\ref{Eq:Bogolubov}) above the
critical point can be obtained from
Eq.(\ref{Eq:Solution-To-Direct-Transformation}) replacing:
$\omega_0\rightarrow\widetilde{\omega}_0-2\zeta$, $g_k\rightarrow
\widetilde{g}_k$.

Equation (\ref{Eq:E-Discrete-Before-QFT}) together with
Eq.(\ref{Eq:E-Discrete-After-QFT}) gives the complete spectrum of
the multimode Dicke model in the quasiclassical limit for both
normal and superradiant phases. Figure 1 illustrates the behavior
of the energy levels before and after the phase transition for a
finite number of modes and $g_k=g \sqrt{\omega_k/2}$, so that
$\gamma=N g^2/2$. The latter choice is for convenience only. One
observes that the excitation energy originating from the atomic
level hits the ground state at the critical point, which results
in the divergence of the spatial parameters, as will be
demonstrated shortly. After the critical point it goes trough the
discrete ``band'' of the field-originated energy levels, creating
an anticrossing structure with each of them. Finally it formes a
detached state developing the ``gap'' above the field levels. The
energy gap for each anticrossing feature is inversely related to
the number of field modes. The energies experience a jump in the
derivatives at the criticality.

It is interesting to note that for large $g_k$, the creation and
annihilation operators $q_N$ and $q_N^\dag$ corresponding to the
separated top energy level become equal to a combination of $a_0$
and $a_0^\dag$ only---namely, $a_0\rightarrow
(3q_N-q_N^\dag)/\sqrt{8}$. At the same time $q_{\alpha<N}$ and
$q_{\alpha<N}^\dag$ converge to $-a_k$ and $-a_k^\dag$, such that
with the corresponding levels they resemble the initial ($g_k=0$)
set of field excitations up to the phase factor. One concludes
that at large $g_k$ the atomic subsystem and field states are
factorized in each eigenstate of the system. The atom-field
interaction merely dresses the atomic ensemble with energy
$\varepsilon \rightarrow 4\gamma$. Therefore, one would expect
zero entanglement between the atoms and the field by that point.
The anticrossing features following the phase transition can,
then, be easily understood in terms of the transfer of an
excitation while the coupling constants are changed adiabatically.
For instance, if one has the top field mode excited initially, the
adiabatic increase of the coupling will result in the transfer of
this excitation to the atomic system with the energy above the
field ``band'' and vice versa; see Fig.~1. Note, however, that all
operators $a_{\alpha<N}$ and $a_{\alpha<N}^\dag$ are shifted after
the critical point.

Below the critical point the field is empty, $\langle a_k^{\dagger
}a_k\rangle =0$, and the atomic system is in its ground state,
$\langle J_z\rangle /j=-1$. Above the critical point the atomic
system loses the magnetization as $\langle J_z\rangle /j=-\gamma
_c/\gamma $ and the field becomes occupied as $\langle
a_k^{\dagger }a_k\rangle =j\omega _0\left( \gamma ^2-\gamma
_c^2\right) /2\gamma \gamma _c$. This is similar to the
single-mode case.\cite{Emary-Brandes}

Let us now proceed to the continuous spectrum with the Ohmic bath
model above criticality. Similarly to
Eq.(\ref{Eq:E-Continuous-Before-QFT}) we find
\begin{equation}\label{Eq:E-Continuous-After-QFT}
\frac{\omega _0^2}{2\Delta} \varepsilon \ln \frac{\Omega
_1+\varepsilon }{\Omega _2+\varepsilon }\frac{ \Omega
_2-\varepsilon }{\Omega _1-\varepsilon }+\varepsilon
^2-16\alpha^2\Delta^2+\omega _0^2 =0.
\end{equation}
The complete energy spectrum of the continuum-field case is
presented in Fig.~2. It consists of two branches located below and
above the energy band. The effective impurity level vanishes at
$\alpha_c$. As the strength of the interaction increases the
dressed level approaches the continuum. At the same time a new
dressed impurity state splits off from the field continuum
developing the gap above the band, $\varepsilon\rightarrow
4\alpha\Delta$. This can be understood comparing the discrete and
continuum spectrum model; cf. Figs. 1 and 2. For larger
bands---i.e., larger $\Delta$---the split-off energy converges to
the limiting linear behavior faster.

Placing the impurity above the field band, one observes similar
behavior; see Fig.~2. The over-band impurity level in this case
diverges from the band starting with $\alpha=0$. The effective
impurity level below the band splits off the field modes and
vanishes at the critical point $\alpha=\alpha_c$. Then it returns
back to the field band. For both positions of the undressed
impurity level the slope of the excitation energy above the band
jumps at the criticality.

\begin{figure}
\includegraphics[width=8cm]{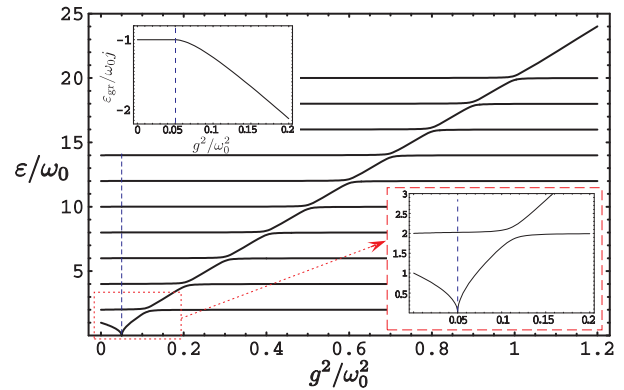}
\caption{Energy spectrum for the discrete model; $N=10$ and
$g_k=g\sqrt{\omega_k/2}$. The top-left inset shows the
ground-state energy of the system. The bottom-right inset presents
the area around the critical point, in which the first excited
state collapses to the ground state. It also magnifies the
anticrossing between the effective impurity level and the first
field mode. In all the frames the vertical dashed line shows the
quantum phase transition point.}
\end{figure}

When the field band is not bounded from above, except for the
gradual effective cutoff, the impurity excitation level placed in
the gap below the field band still undergoes the transition at the
critical point, which is defined as $\alpha
_c=\omega_0/4\Delta_\textrm{eff}$, and then merges with the field
band. The split-off level above the band is not developed in this
case.

%
\textit{Critical exponents.} An important property of phase
transitions is that systems with different microscopic dynamics
behave equivalently at criticality. Their behavior depends only on
the dimension of the system and the symmetry of the order
parameter.\cite{Sachdev} This phenomenon is known as universality.
The parameters describing this critical behavior are universal
quantities. Their behavior at the criticality is completely
described in terms of a critical exponent.

Let us analyze the behavior of the lowest excited state energy
(relatively to the ground state) in the vicinity of the critical
point. We consider the discrete case first, factoring the
magnitude of the interaction for convenience as $g_k\equiv
g\,\chi_k$. Here $g$ is varied in the vicinity of the critical
value $g_c=\sqrt{\omega_0}/\sqrt{4\sum_k \chi_k^2/\omega_k}$. This
is feasible as far as the shape of the dependence stays the same.
Stepping away form the critical point infinitesimally as
$g=g_c\pm\delta g$, where the negative sign is taken for
Eq.(\ref{Eq:E-Discrete-Before-QFT}) and the positive for
Eq.(\ref{Eq:E-Discrete-After-QFT}), one obtains
\begin{equation}
\varepsilon\sim N_c\left|g -g_c\right|^{z\nu}.
\end{equation}
Here $z\nu=1/2$ is the critical exponent. The nonuniversal
constant of proportionality which defines the energy scale from
the left of the critical point is
\begin{equation} \label{Eq:N-constant-before}
N_c=\left(\frac{2\omega _0^2}{g_c+4\omega _0g_c^3\sum_{k=1}^N\chi
_k^2\omega _k^{-3}}\right)^{1/2}.
\end{equation}
After the critical point it is greater by a factor of $\sqrt{2}$.
This behavior marks the second-order quantum phase transition.

In the case of continuously distributed field modes with the Ohmic
model one varies $\alpha$ around the critical value $\alpha
_c=\omega_0/4\Delta $ in Eq.(\ref{Eq:E-Continuous-Before-QFT}) and
(\ref{Eq:E-Continuous-After-QFT}), observing similar behavior for
the impurity excitation energy
\begin{equation}
\varepsilon \sim N_c^\prime\left| \alpha -\alpha _c\right|
^{z\nu},
\end{equation}
with the same critical exponents $z=2$ and $\nu=1/4$. The
nonuniversal coefficient $N_c^\prime$ to the left of the critical
point is
\begin{equation}
N_c^{\prime }=\sqrt{\frac{4\omega _0\Delta }{1+\frac{\omega
_0^2}{\Omega _1\Omega _2}}}\,.
\end{equation}
After the critical point it acquires a factor of $\sqrt{2}$.

\begin{figure}[tbp]
\includegraphics[width=8cm]{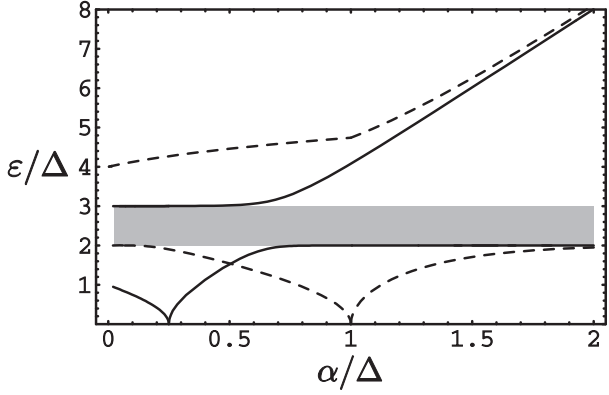}
\caption{Energy spectrum for the continuous field distribution.
The field band is shown by the shaded area. The solid curve
represents the effective impurity excitation energy. It begins
below the band at $\alpha=0$ and ends above the band at
$\alpha\rightarrow\infty$ as purely atomic excitation energy. The
quantum phase transition occurs when the first excitation energy
of the system (solid bottom curve) hits the ground state at
$\alpha_c=\Delta/4$. The ground-state energy changes similarly to
the discrete case and is not shown here. The parameters are
$\omega_0=\Delta$, $\Omega_1=2\Delta$, and $\Omega_2=3\Delta$. The
dashed top curve shows the effective impurity when it is placed at
$\omega_0=4\Delta$. In this case the quantum phase transition is
manifested by the first exited level (bottom dashed curve) as it
splits off below the band and hits the ground-state energy at
$\alpha_c=\Delta$.}
\end{figure}
%
%
\textit{Entanglement as an order parameter.}The quantum phase
transition is governed by a divergent correlation length at the
critical point. This is similar to what happens in the classical
case. In the quantum case, though, there are correlations which do
not have a classical analog. These purely quantum correlations are
known as entanglement. It was
demonstrated\cite{Osterloh,Osborne,Wu} that there is a close
relation between quantum phase transitions and entanglement. For a
single-mode model it was shown\cite{Emary-QPT-Concurrence} that
the atoms become strongly entangled in the vicinity of the
critical point. Osterloh \textit{et al.}\cite{Osterloh} have found
that the concurrence is universal for the \textit{XY} models. In
what follows, we investigate the entanglement between the impurity
atoms in the multimode field model and analyze how it is affected
by the number of modes involved.

The entanglement of an effectively bipartite system can be
measured by the concurrence,\cite{Concurrence} which is a convex
function of the entanglement of formation.\cite{Bennett} For a
given two-atom density matrix it is defined\cite{Concurrence} as
\begin{equation}
C(\rho )=\max \Big\{ 0,2\max_j \lambda _i
-\sum\limits_{j=1}^4\lambda _j\Big\},
\end{equation}
where $\lambda_j$ are the eigenvalues of the Hermitian matrix
$R=\sqrt{\sqrt{\rho}\widetilde{\rho}\sqrt{\rho}}$ and
$\widetilde{\rho}=\left( \sigma _{y}\otimes \sigma _{y}\right)
\rho ^{*}\left( \sigma _{y}\otimes \sigma _{y}\right)$. Due to the
symmetry of the atomic system with respect to the exchange of
atoms, one can evaluate pairwise concurrence, extracting
two-particle states from the multiparticle symmetric states. Wang
and M{\o}lmer have demonstrated\cite{Wang-Molmer} that the
two-atom reduced density matrix written in the standard basis
$\big\{\left| \uparrow \uparrow \right\rangle ,\left| \uparrow
\downarrow \right\rangle ,\left| \downarrow \uparrow \right\rangle
,\left| \downarrow \downarrow \right\rangle\big\}$ takes the form
\begin{equation}\label{Eq:Bipartite-Density-Matrix}
\rho_{12} =\left(
\begin{array}{cccc}
v_+ & x_+^*& x_+^*& u^*  \\
x_+ & w    & y^*  & x_-^*\\
x_+ & y    & w    & x_-^*\\
u   & x_-  & x_-  & v_-
\end{array}
\right).
\end{equation}
As mentioned earlier, utilizing the symmetry of the atomic state
under the exchange of particles one can express all the elements
in Eq.(\ref{Eq:Bipartite-Density-Matrix})---i.e., the expectation
values $\langle\sigma_{1i}\sigma_{2j}\rangle$, etc.---via the
expectation values of the collective spin
operators,\cite{Wang-Molmer}
\begin{eqnarray}\nonumber \label{Eq:Density-Matrix-Elements}
v_{\pm }\!\!\!&=&\!\!\frac{M^2-2M+4\left\langle J_z^2\right\rangle
\pm 4\left\langle J_z\right\rangle \left( M-1\right) }{4M\left(
M-1\right) },
\\\nonumber
x_{\pm }\!\!\!&=&\!\!\frac{\left( M-1\right) \left\langle
J_{+}\right\rangle \pm \left\langle \left[ J_{+},J_z\right]
_{+}\right\rangle }{2M\left( M-1\right) },\,\,\,\,\,\,
u\!=\!\frac{\left\langle J_{+}^2\right\rangle }{M\left(
M-1\right)}
\\
w\!\!\!&=&\!\!\frac{M^2-4\left\langle J_z^2\right\rangle
}{4M\left( M-1\right) },\,\,\,\,\,\, y=\frac{2\left\langle
J_x^2+J_y^2\right\rangle -M}{2M\left( M-1\right) }.
\end{eqnarray}

To calculate the above expectation values below and above the
phase transition, we utilize Eq.(\ref{Eq:HP-relations}) in the
limit of large $M$. Below the critical point we obtain
\begin{eqnarray}\label{Eq:Averages-of-Js-Before}\nonumber
\left\langle J_z^2\right\rangle &=&\frac{M^2}{4}-M\left\langle
b^{\dagger }b\right\rangle+\left\langle b^{\dagger }bb^{\dagger
}b\right\rangle+{\cal O}(M^{-s}),
\\\nonumber
\left\langle
J_{+}^2\right\rangle\!\! &=&\!\!\big( M-\sqrt{M}\big)\left\langle
b^{\dagger }b^{\dagger }\right\rangle+{\cal
O}(M^{-s}),
\\
\left\langle J_z\right\rangle\!\!&=&\!\! -\frac M 2+\left\langle
b^{\dagger }b\right\rangle+{\cal O}(M^{-s}), \,\,\,\,s\geq1/2.
\end{eqnarray}
The other expectation values are either trivial ($=0$) or can be
derived from Eq.(\ref{Eq:Averages-of-Js-Before}). After the phase
transition we have
\begin{eqnarray}\label{Eq:Averages-of-Js-After}
\nonumber \left\langle J_z^2\right\rangle^\prime
\!\!&=&\!\!\frac{M^2}{4}\eta ^2+\frac M 2\left( 1-\eta
\right)+M\left( 1-\eta \right) \left\langle b^{\dagger
2}\right\rangle^\prime
\\\nonumber
&+&M\left( 1-2\eta \right) \left\langle b^{\dagger
}b\right\rangle^\prime+\left\langle b^{\dagger }bb^{\dagger
}b\right\rangle^\prime +{\cal O}(M^{-s}),
\\\nonumber
\left\langle J_{+}^2\right\rangle^\prime
\!\!&=&\!\!\frac{M^2}{4}\left( 1-\eta ^2\right)+\frac M 4\left(
\eta -1\right)+\frac M 2\left( 3\eta -1\right) \left\langle
b^{\dagger 2}\right\rangle^\prime
\\\nonumber
&+&\frac {3M}2\left(
\eta -1\right) \left\langle
b^{\dagger }b\right\rangle^\prime +{\cal O}(M^{-s}),
\\
\left\langle J_z\right\rangle^\prime\!\!&=&\!\!-\frac M
2\eta+\left\langle b^{\dagger }b\right\rangle^\prime +{\cal
O}(M^{-s}),\,\,\,\,s\geq1/2.
\end{eqnarray}
We point out that the ground states in Eqs.
(\ref{Eq:Averages-of-Js-Before}) and
(\ref{Eq:Averages-of-Js-After}) are different, which is the reason
for the ``primes'' in Eqs.(\ref{Eq:Averages-of-Js-After}). Note
also that after the critical point one has to include the shift of
the bosonic operators in Eqs.(\ref{Eq:HP-relations}) to obtain
Eqs.(\ref{Eq:Averages-of-Js-After}).

The expectation values of the bosonic operators before the
critical point are found utilizing the transformation
(\ref{Eq:Bogolubov}) as $\langle b^{\dagger }b\rangle =\sum_\beta
\overline{A}_{0\beta }^2$, $\langle b^{\dagger 2}\rangle
=\sum_\beta A_{0\beta }\overline{A}_{0\beta }$, and $\langle
b^{\dagger }bb^{\dagger }b\rangle =\sum_{\beta ,\gamma }A_{0\beta
}\overline{A}_{0\beta }A_{0\gamma } \overline{A}_{0\gamma
}+\sum_{\beta ,\gamma }\overline{A}_{0\beta }^2A_{0\gamma
}^2+\sum_{\beta ,\gamma }\overline{A}_{0\beta }^2\overline{A}
_{0\gamma }^2$. Here we omit the complex conjugation since all the
amplitudes are real. Together with Eqs.
(\ref{Eq:Solution-To-Direct-Transformation}) and
(\ref{Eq:E-Discrete-Before-QFT}) these relations give the
expectation values of the total spin operators. A similar
procedure leads to the explicit expressions for the expectation
values after the critical point (\ref{Eq:Averages-of-Js-After}).
\begin{figure}[tbp]
\includegraphics[width=8cm]{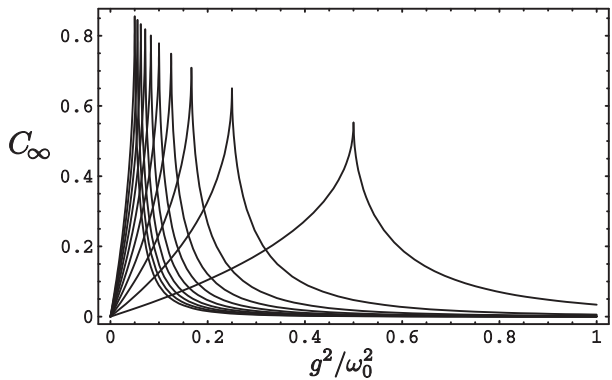}
\caption{Concurrence for the discrete model with different numbers
of field modes. From right to left, $N=1,2,\ldots,10$. The other
parameters are the same as in Fig.~1.}
\end{figure}

For the Dicke model the reduced density matrix
(\ref{Eq:Bipartite-Density-Matrix}) simplifies,
$x_+=x_+^*=x_-=x_-^*=0$, $y=y^*=w$, and the eigenvalues of the
matrix $R$ are $\lambda _1=2w$,
$\lambda_{2,3}=\big|u\pm\sqrt{v_{+}v_{-}}\big|$, and $\lambda
_4=0$. One can verify that $\lambda _2>\lambda _3>\lambda
_4>\lambda _1$, $u<\sqrt{v_{+}v_{-}}$, and
$2w<u+\sqrt{v_{+}v_{-}}$. As a result the concurrence below and
above the phase transition has the form $C=2(u-w)$. However, in
the thermodynamic limit $M\rightarrow\infty$, the pairwise
concurrence vanishes as $C={\cal O}(M^{-1})$ due to exchange
symmetry, so we define the scaled concurrence
\cite{Emary-QPT-Concurrence, VidalConcurrence, VidalCollective}
\begin{equation}
C_M(\rho )=2M(u-w).
\end{equation}
Using relations (\ref{Eq:Density-Matrix-Elements}),
(\ref{Eq:Averages-of-Js-Before}), and
(\ref{Eq:Averages-of-Js-After}) and the expressions for $\langle
b^{\dagger }b\rangle$, $\langle b^{\dagger 2}\rangle$, and
$\langle b^{\dagger }bb^{\dagger }b\rangle$ we obtain the limiting
value of the scaled concurrence,
$C_{\infty}=\lim_{M\rightarrow\infty}C_M(\rho )$, in the form
\begin{equation}\label{Eq:Concurrence-in-the-Limit}
C_\infty = \left(1+\theta \right) \sum_\beta W_\beta \frac{\omega
_0-\varepsilon _\beta }{2\omega _0}+\frac 12\left( 1-\theta
\right)\,,
\end{equation}
where
\begin{equation}\label{Eq:Wk-coefficient}
W_\beta =\left[ 1+\sum_k\frac{4g_k^2\omega _0\omega _k}{\left(
\omega _k^2-\varepsilon _\beta ^2\right) ^2}\right] ^{-1}.
\end{equation}
Here $\theta\equiv1$ below the critical point, $\gamma<\gamma_c$,
and $\theta\equiv\eta$ above the critical point,
$\gamma>\gamma_c$. Also we imply the replacements $\omega
_0\rightarrow \widetilde{\omega } _0-2\zeta $, $g_k\rightarrow
\widetilde{g}_k$ for $\gamma>\gamma_c$. In Fig.~3 we plot
$C_\infty$ as a function of the coupling strength for various $N$,
using the coupling constants in the form $g_k=g\sqrt{\omega_k/2}$
for an illustrative example. The concurrence reaches the maximum
and breaks at the critical point. We note that the increase in the
number of the field modes leads to the corresponding increase of
the maximum bipartite entanglement in the atomic system.

Let us investigate the critical behavior of the concurrence.
Introducing an infinitesimal deviation of the coupling from the
critical point $g=g_c\pm\delta g$ one obtains $\delta
C_\infty=-W_0^c\delta\varepsilon/\omega_0$ from the first term in
Eq.(\ref{Eq:Concurrence-in-the-Limit}). Here $W_0^c$ is $W_0$,
[see Eq.(\ref{Eq:Wk-coefficient})], evaluated at the critical
point. As a result one has
\begin{equation}\label{Eq:Concurrence-Scaling}
C_\infty \sim -\frac 1{\omega _0}W_0^cN_c\left| g-g_c\right|
^{z\nu },
\end{equation}
with the critical exponent $z\nu=1/2$, which defines entanglement
(via concurrence) as an order parameter. One can easily generalize
these results to the case of the continuum field model.

%
{\it $A^2$-term.} In many circumstances the overall Hamiltonian of
the ensemble of two-state systems interacting with electromagnetic
field would involve an additional field term.\cite{note1} One
would expect a Hamiltonian of the form $H(t) = \sum_{j = 1}^M
[\mathbf{p}_j-\frac{e}{c}\mathbf{A}(t)]^2/2m  + \sum_{j = 1}^M
V_j(\mathbf{r}_j)$, where $\mathbf{p}_j$ and $\mathbf{r}_j$ are
momentum and coordinate operators of the $j$th subsystem with
confinement potential $V_j$. Rewriting the Hamiltonian in the
eigenspace of the two-state systems with the assumptions discussed
above we arrive at Eq.(\ref{Eq:H}) with an additional term
$M\frac{e^2}{2mc^2} \mathbf{A}^2$. In terms of the standard
bosonic operators it is $\sum_k s_k (a_k^\dag + a_k)^2$ where
$s_k$ depends on the density of states and therefore is finite
when $M \to \infty$. Often this term is small with respect to the
linear one and therefore is not considered. Nevertheless, it was
shown \cite{Rzazewski,Rzazewski2,Thompson1,Thompson2} that it can
lead to the absence of the classical phase transition. Below we
demonstrate that the quantum phase transition in the Dicke model
with the $A^2$ term is present and can still be qualitatively
described by the above results.

We consider a single-mode model with $\omega_0=\omega_k=\omega$.
The diagonalization of Eq.(\ref{Eq:H}) with the $A^2$ term---i.e.,
$s(a^\dag + a)^2$---yields an expression for the spectrum in the
form $\varepsilon^2 = 1 + 2s \pm \sqrt {(1+2s)^2 - (1+4s-4g^2)}$.
Here $\varepsilon$, $s$, and $g$ are in units of $\omega$. Exactly
as before, one finds the critical point at $g_c = \sqrt{1+4s}/2$,
where the energy gap between the first excited and the ground
state energy levels vanishes. Above this point the symmetry
breaks. At the criticality the energy scales as $\varepsilon \to
\sqrt{{4g_c}/(1 + 2s)}|g-g_c|^{1/2}$. After the critical point
there is additional factor of $\sqrt 2$ on the right-hand side. As
a result, one observes the same transition with the same critical
exponents and corrected position of the critical point.

In the multimode case the derivation is more cumbersome due to the
cross terms in $\mathbf{A}^2$. Considering the above strategy it
is expected that the generalization will yield similar unimportant
corrections. We leave a thorough proof of this statement to be
given elsewhere.

In summary, we investigated the model of the atomic ensemble
interacting with an arbitrary discrete or continuous set of
bosonic modes. The system was shown to undergo a quantum phase
transition of the second order. In addition a series of
anticrossing features following the criticality was revealed in
the field band. Critical behavior of the parameters was found
explicitly. It was demonstrated that the pairwise concurrence of
the atomic system is broken at criticality with the same critical
exponent as the energy.

{\acknowledgements The authors acknowledge contributing
discussions with V. Privman and D. Mozyrsky. D.T. acknowledges the
hospitability of the Center for Nonlinear Studies at Los Alamos
National Laboratory. The work was supported in part by the NSF
under Grant No. DMR-0121146.


\begin{thebibliography}{99}
\bibitem{note1} The corrections for phonon fields may be more complicated.
\end{thebibliography}
\end{document}